\newcommand{\fethirteen}{\mathrm{Fe^{13+}}}
\newcommand{\fesixteen}{\mathrm{Fe^{16+}}}
\newcommand{\feseventeen}{\mathrm{Fe^{17+}}}
\documentclass[manuscript]{aastex}
\usepackage{amsbsy}
\usepackage{amsfonts}
\usepackage{amssymb}

\usepackage{graphicx}
\usepackage{graphics}
\usepackage{bm}

\begin{document}


\title{Storage Ring Cross Section Measurements for Electron Impact Single and Double Ionization of $\mathrm{\mathbf{Fe^{13+}}}$ and Single Ionization of $\mathrm{\mathbf{Fe^{16+}}}$ and $\mathrm{\mathbf{Fe^{17+}}}$}

\author{M. Hahn\altaffilmark{1}, A. Becker\altaffilmark{2}, D. Bernhardt\altaffilmark{3}, M. Grieser\altaffilmark{2}, C. Krantz\altaffilmark{2},  M. Lestinsky\altaffilmark{4}, A. M\"{u}ller\altaffilmark{3}, O. Novotn\'{y}\altaffilmark{1}, R. Repnow\altaffilmark{2}, S. Schippers\altaffilmark{3}, K. Spruck\altaffilmark{3}, A. Wolf\altaffilmark{2}, and D. W. Savin\altaffilmark{1}}

\altaffiltext{1}{Columbia Astrophysics Laboratory, Columbia University, 550 West 120th Street, New York, NY 10027 USA}
\altaffiltext{2}{Max-Planck-Institut f\"{u}r Kernphysik, Saupfercheckweg 1, 69117 Heidelberg, Germany}
\altaffiltext{3}{Institut f\"{u}r Atom- und Molek\"{u}lphysik, Justus-Liebig-Universit\"{a}t Giessen, Leihgesterner Weg 217, 35392 Giessen, Germany}
\altaffiltext{4}{GSI Helmholtzzentrum f\"ur Schwerionenforschung, Planckstr. 1, 64291 Darmstadt, Germany }

\date{\today}
\begin{abstract}
	
	We report measurements of electron impact ionization (EII) for $\fethirteen$, $\fesixteen$, and $\feseventeen$ over collision energies from below threshold to above 3000~eV. The ions were recirculated using an ion storage ring. Data were collected after a sufficiently long time that essentially all the ions had relaxed radiatively to their ground state before data were collected. For single ionization of $\fethirteen$ we find that previous single pass experiments are more than 40\% larger than our results. Compared to our work, the theoretical cross section recommended by \citet{Arnaud:ApJ:1992} is more than 30\% larger, while that of \citet{Dere:AA:2007} is about $20\%$ greater. Much of the discrepancy with \citet{Dere:AA:2007} is due to the theory overestimating the contribution of excitation-autoionization via $n=2$ excitations. Double ionization of $\fethirteen$ is dominated by direct ionization of an inner shell electron accompanied by autoionization of a second electron. Our results for single ionization of $\fesixteen$ and $\feseventeen$ agree with theoretical calculations to within the experimental uncertainties. 

\end{abstract}
	
\maketitle
	
\section{Introduction}
	
	Spectroscopic diagnostics of cosmic sources rely on accurate charge state distribution (CSD) calculations \citep{Brickhouse:AIP:1996, Landi:AA:1999}. 
In stellar coronae, supernova remnants, galaxies, the intracluster medium of galaxy clusters, and other collisionally ionized plasmas, the balance between electron impact excitation (EII) and electron-ion recombination determines the CSD \citep{Bryans:ApJ:2009}. Thus, accurate EII data are needed to derive the CSD. For most objects, the temperature varies slowly enough that only electron impact single ionization (EISI) matters \citep{Tendler:PhysLett:1984}; but when there is rapid heating, electron impact double ionization (EIDI) can be important \citep{Muller:PhysLett:1986}. 
	
	Most EII data come from theoretical calculations, as it is not possible to measure ionization for every ion. Experiments serve to benchmark these theoretical calculations. However, a major limitation of most existing EII measurements is that the ion beams used contained an unknown population of metastable ions. As the cross section for ionization from a metastable level generally differs from that for the ground level, the results of such experiments can be ambiguous and do not provide a clear test for theory.
		
	Here we report EII measurements for Al-like $\fethirteen$, Ne-like $\fesixteen$, and F-like $\feseventeen$. Of these three ion species, previous measurements exist only for $\fethirteen$ \citep{Gregory:PRA:1987}. However, those measurements were performed using a crossed beams experiment and suffer from an unknown metastable fraction. We used an ion storage ring to avoid this problem. The ions were recirculated in the ring for several seconds before collecting data. This allowed any metastable levels to relax radiatively to the ground state. Our data thereby provide an unambiguous test for theoretical models. 
	
	The EISI cross section for $\fethirteen$ was measured from about 300~eV to 3100~eV, where the direct ionization channels are
\begin{equation}
\mathrm{e^{-} + Fe^{13+}} (2s^2\, 2p^6\, 3s^2\, 3p) \rightarrow \left\{ \begin{array}{l} 
													\mathrm{Fe^{14+}} (2s^2\, 2p^6\, 3s^2) + 2\mathrm{e^{-}} \\
													\mathrm{Fe^{14+}} (2s^2\, 2p^6\, 3s\, 3p) + 2\mathrm{e^{-}} \\
													\mathrm{Fe^{14+}} (2s^2\, 2p^5\, 3s^2\, 3p) + 2\mathrm{e^{-}} \\
													\mathrm{Fe^{14+}} (2s\, 2p^6\, 3s^2\, 3p) + 2\mathrm{e^{-}} 
													\end{array} \right. .
\label{eq:transitions13}
\end{equation}	
The corresponding thresholds for $3p$ and $3s$ ionization are 392.2~eV and 421.2~eV, respectively \citep{NIST:2012}. Direct ionization of a $2p$ electron is possible above $1127.3$~eV \citep{NIST:2012} and ionization of the $2s$ is possible above about $1270$~eV \citep{Kaastra:AAS:1993}. However, following direct ionization of a principal quantum number $n=2$ electron, the resulting excited state is expected to stabilize via autoionization with a probability of greater than 90\%. This produces double ionization rather than single ionization \citep{Kaastra:AAS:1993}. Ionization can also occur through Excitation-autoionization (EA). For example, excitation of a $3s$ electron to a doubly excited level lying in the continuum is possible starting from the 392.16~eV ionization threshold. The system can then autoionize resulting in EISI. EA via $n=2$ to $n=3$ excitations are predicted to occur starting at $\sim 700$~eV and via $2\rightarrow4$ $\sim 900$~eV \citep{Pindzola:PRA:1986,Arnaud:ApJ:1992,Dere:AA:2007}. 

	The EIDI cross section for $\fethirteen$ forming Fe$^{15+}$ was investigated from below just the threshold for direct double ionization at 848.4~eV up to 3100~eV. For highly charged ions the dominant double ionization process is expected to be single ionization of an electron in the $n=2$ level forming a state that stabilizes by emission of a second electron \citep{Muller:PRL:1980}. The threshold for this ionization-autoionization process is $1127.3$~eV \citep{NIST:2012}.
	
	The EISI cross sections for $\fesixteen$ and $\feseventeen$ were measured from $1100$~eV to $3200$~eV, where the direct ionization channels are
\begin{equation}
\mathrm{e^{-} + Fe^{16+}} (2s^2\, 2p^6) \rightarrow \left\{ \begin{array}{l} 
													\mathrm{Fe^{17+}} (2s^2\, 2p^5) + 2\mathrm{e^{-}} \\
													\mathrm{Fe^{17+}} (2s\, 2p^6) + 2\mathrm{e^{-}} 
													\end{array} \right. .
\label{eq:transitions16}
\end{equation}	
The ionization thresholds for these channels are 1262.7~eV for ionization of a $2p$ electron and 1394.7~eV for ionization of a $2s$ electron \citep{NIST:2012}. EA may occur through excitation of the $2s$ electron beginning from the ionization threshold of 1262.7~eV. 
Similarly, for $\feseventeen$ the direct ionization channels are
\begin{equation}
\mathrm{e^{-} + Fe^{17+}} (2s^2\, 2p^5) \rightarrow \left\{ \begin{array}{l} 
													\mathrm{Fe^{18+}} (2s^2\, 2p^4) + 2\mathrm{e^{-}} \\
													\mathrm{Fe^{18+}} (2s\, 2p^5) + 2\mathrm{e^{-}} 
													\end{array} \right. .
\label{eq:transitions17}
\end{equation}	
These have ionization thresholds of 1357.8 for the $2p$ electron and 1472.2 for the $2s$ electron. Here again EA from excitation of a $2s$ electron is possible beginning at the ionization threshold of 1357.8~eV. 

\section{Experimental Method and Analysis}\label{sec:exan}
	
	Cross section measurements were performed using the TSR heavy ion storage ring. This facility is located at the Max-Planck-Institut f\"{u}r Kernphysik in Heidelberg, Germany. The procedures used here are basically the same as those described by \citet{Linkemann:PRL:1995} and \citet{Hahn:ApJ:2010, Hahn:ApJ:2011, Hahn:ApJ:2011a, Hahn:PRA:2012, Hahn:ApJ:2012}. Below we outline the method and provide some details pertinent to the present measurements. 
	
	First a beam of iron ions was introduced into TSR. The ion beam energies were $141.0$~MeV for $\fethirteen$, $176.9$~MeV for $\fesixteen$ and $183.8$~MeV for $\feseventeen$. In each case the isotope $^{56}$Fe was used for the experiment. Two electron beams in the ring, dubbed the Cooler and the Target, where merged with the ions. Each electron beam is located in a different section of the ring. Initially the electron beams were used to cool the ion beam. That is, the energy of both electron beams was fixed to one where the electron velocity closely matched the average ion velocity, allowing elastic electron-ion collisions to reduce the energy spread of the ion beam \citep{Poth:PhysRep:1990}. This initial cooling period lasted three seconds. 
	
	During cooling, metastable states in the ion beam radiatively decayed. For $\fethirteen$ there are two metastable levels with relatively long lifetimes. These are the $3s^2\,3p\,^{2}P_{3/2}$ level, whose decay to the ground state forms the well-known coronal green line \citep{Edlen:ZA:1943, Esser:JGR:1995}, and the $3s\,3p\,3d\,^{4}F_{9/2}$ level. The $3s^2\,3p\,^{2}P_{3/2}$ lifetime has been measured experimentally to be about 16.73~ms \citep{Beiersdorfer:ApJ:2003, Brenner:PRA:2007}. The lifetime of the $3s\,3p\,3d\,^{4}F_{9/2}$ level has been calculated theoretically to be 17.7~ms \citep{Trabert:PhysScr:1993}. The similar lifetimes of these levels have been a source of systematic uncertainty in lifetime measurements at TSR \citep{Trabert:JPhysB:2002, Trabert:JPhysB:2009, Trabert:JPhysB:2010}. 
For $\fesixteen$ the longest lived metastable levels are the $2s^2\,2p^5\,3s\,^{3}P_{0,2}$, which have predicted lifetimes of $63.3$~$\mathrm{\mu s}$ and $4.5$~$\mathrm{\mu s}$, respectively \citep{Liang:AA:2010, Landi:ApJ:2012}. The longest lived $\feseventeen$ metastable level is the $2s^2\,2p^5\,^{2}P_{1/2}$ level, which has a predicted lifetime of 51.5~$\mathrm{\mu s}$ \citep{DelZanna:AA:2006, Landi:ApJ:2012}. Since all these lifetimes are much shorter than the $3$~s cooling time, the metastable population for each ion is expected to be negligible during measurement. 
	
	After cooling, the Target was maintained at the cooling energy while the Cooler electron beam energy was varied so as to enable electron-ion collision studies at different energies. Ionized products from collisions in the Cooler section were diverted by a downstream dipole magnet onto a particle detector. The measurement energy was stepped through a range of energies. In between each measurement step the ionization count rate was recorded for a fixed reference energy. This allowed us to assess the rate for background stripping off the residual gas. Ideally, this reference rate should be measured below the EII threshold, but at high measurement energies we were limited by the dynamic range of the electron beam power supply. Hence, for these energies the reference energy was set to a point where the ionization cross section was already measured in lower energy scans, allowing the background rate to be derived. The energy range for which the reference point was set below threshold was $E \leq 1050$~eV for $\fethirteen$ EISI and $E \leq 1500$~eV for EIDI, $E \leq 1980$~eV for $\fesixteen$ EISI, and $E \leq 1950$~eV for $\feseventeen$ EISI. 
	
	The EII cross section $\sigma_{\mathrm{I}}$ was obtained from the difference between the measured count rate and the background signal, normalized by the stored ion number and the electron density \citep{Hahn:ApJ:2011}. The uncertainty introduced by the detector efficiency is about 3\% \citep{Rinn:RSI:1982}. Here and throughout all uncertainties are given at an estimated $1\sigma$ level. The electron density has an uncertainty of about 3\% \citep{Lestinsky:ApJ:2009}. The stored ion number was derived from the ion current measured with a beam profile monitor \citep[BPM;][]{Hochadel:NIMA:1994}. We calibrated the BPM several times during the measurement by comparing with the ion current measurement from a DC transformer \citep{Unser:IEEE:1981}. The calibration was performed using currents of up to 21~$\mathrm{\mu A}$ \citep{Hahn:ApJ:2011a}. However, the DC transformer is not sensitive to the $0.5$ - $5$~$\mathrm{\mu A}$ currents present during measurement and could not be used directly for the analysis. We estimate that the uncertainty of the BPM contributes 15\% to the experimental systematic uncertainty.
	
	Energy dependent pressure fluctuations change the background rate and can systematically distort the measured cross section. We corrected for these following \citet{Hahn:ApJ:2010}. The magnitude of the correction was $(0.8 \pm 0.2)\%$ for $\fethirteen$ EISI, $(4\pm2)\%$ for $\fesixteen$ and $(5\pm2)\%$ for $\feseventeen$. For the $\fethirteen$ EIDI measurements we did not find any systematic pressure fluctuations and so no correction was necessary. Because this correction could only be used when the reference point was below the threshold for ionization, there are other uncertainties on the cross sections in the higher energy ranges. The experimental uncertainties are given in Table~\ref{table:err}. 
	
\section{Results and Discussion}\label{sec:res}

\subsection{Single Ionization} \label{subsec:single}

\subsubsection{Cross Sections}\label{subsubsec:cross}

\begin{figure}
\centering \includegraphics[width=0.9\textwidth]{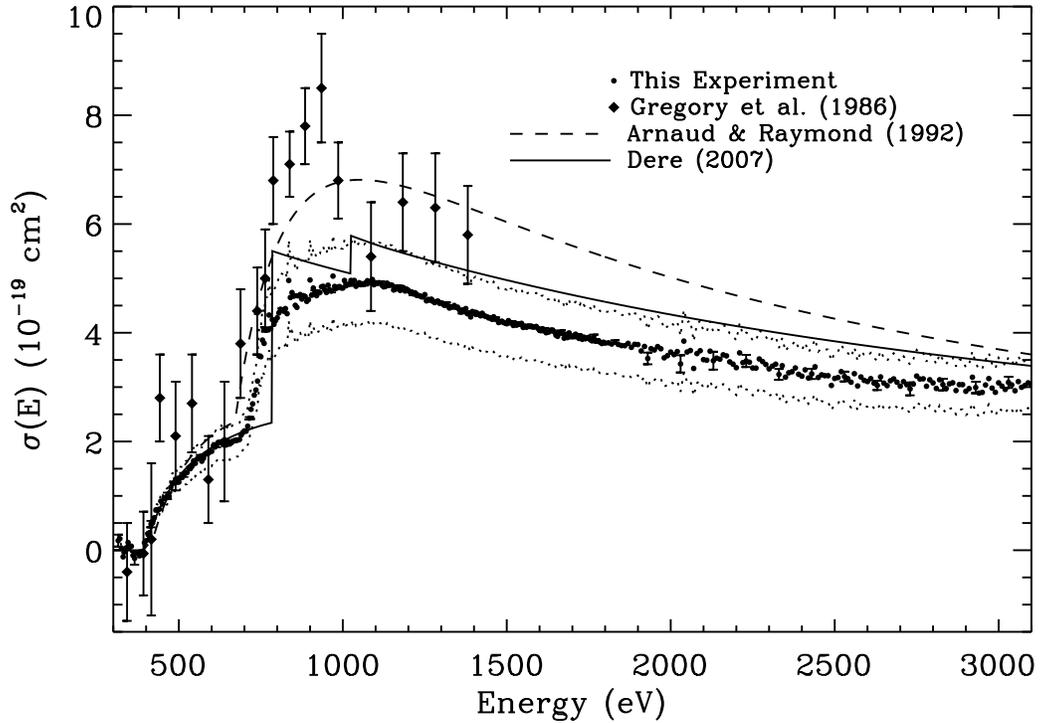}
\caption{\label{fig:fe13cross} EISI cross section for $\fethirteen$ forming Fe$^{14+}$ (circles). The dotted curves illustrate the $1\sigma$ systematic uncertainties. Statistical uncertainties are indicated by the error bars on selected points, but in many cases they are smaller than the symbol size. The experimental results of \citet{Gregory:PRA:1987} are shown by the diamonds. The dashed and solid curves show the theoretical cross sections given by \citet{Arnaud:ApJ:1992} and \citet{Dere:AA:2007}, respectively.
}
\end{figure}

	Figure~\ref{fig:fe13cross} shows the EISI cross section for $\fethirteen$ forming Fe$^{14+}$. These data are also available in the electronic edition of this journal as a table following the format of Table~\ref{table:fe13cross}. In Figure~\ref{fig:fe13cross} the filled circles show the measured cross section and the dotted curves illustrate the $1\sigma$ systematic uncertainty. Error bars on selected points represent the $1\sigma$ errors due to counting statistics. In some cases the error bars are smaller than the symbol size because the magnitude of the statistical uncertainty varies from about 1\% to 4\%, being smaller in places where more data were collected. 
	
	The diamonds in Figure~\ref{fig:fe13cross} show crossed beams EII measurements of \citet{Gregory:PRA:1987}. These results agree with our measurements from threshold to about 700~eV, but they are about $40\%$ larger above 700~eV. This discrepancy is well outside the uncertainties of their and our measurements and is likely due to metastable ions in the crossed beams experiment. Although \citet{Gregory:PRA:1987} do not discuss the possible influence of metastables for this particular ion, the lifetimes of the $\fethirteen$ metastable levels suggest that they could have been present. Their experiment used a 10~kV ion beam. Given that their device had a length scale of several meters, this implies that metastables with lifetimes $\gtrsim10$~$\mathrm{\mu s}$ could remain in the beam. As discussed above, the $\fethirteen$ metastable levels $3s^2\,3p\,^{2}P_{3/2}$ and $3s\,3p\,3d\,^{4}F_{9/2}$ have lifetimes of $\approx16.7$~ms and $\approx 17.7$~ms, respectively, and could therefore both be present in the crossed beams experiment. The ionization threshold for the $3s\,3p\,3d\,^{4}F_{9/2}$ level lies 81.86~eV below the ground state ionization threshold \citep{NIST:2012}. However, the \citet{Gregory:PRA:1987} cross section does not show any contribution below the ground state threshold. This apparently low abundance of the $3s\,3p\,3d\,^{4}F_{9/2}$ level may be due to its relatively high excitation energy and to not being strongly populated by cascades. 
It appears that it is primarily the $3s^2\,3p\,^{2}P_{3/2}$ level that is present in their beam, in addition to the ground level. This is also consistent with the observation that our results and those of \citet{Gregory:PRA:1987} agree at energies where direct ionization dominates, but disagree above the $2\rightarrow3$~EA threshold. The $3s^2\,3p\,^{2}P_{3/2}$ is part of the ground term and so it is expected to have a direct ionization cross section similar to that of the ground state, while the autoionization probabilities of levels above the $2\rightarrow3$ EA threshold could be different for the $^{2}P_{3/2}$ state.   
	
	For comparison, Figure~\ref{fig:fe13cross} illustrates the recommended cross section of \citet{Arnaud:ApJ:1992}, which is based on the theoretical work of \citet{Younger:JQSRT:1983} and \citet{Pindzola:PRA:1986}. This cross section is up to 35\% larger than our results. The reason for this is not clear. Also shown is the distorted wave results of \citet{Dere:AA:2007}, which agrees with the measurement to within the level of the experimental uncertainties, though significant structural differences remain between the results of \citet{Dere:AA:2007} and ours, as discussed below. 
	
\begin{figure}
\centering \includegraphics[width=0.9\textwidth]{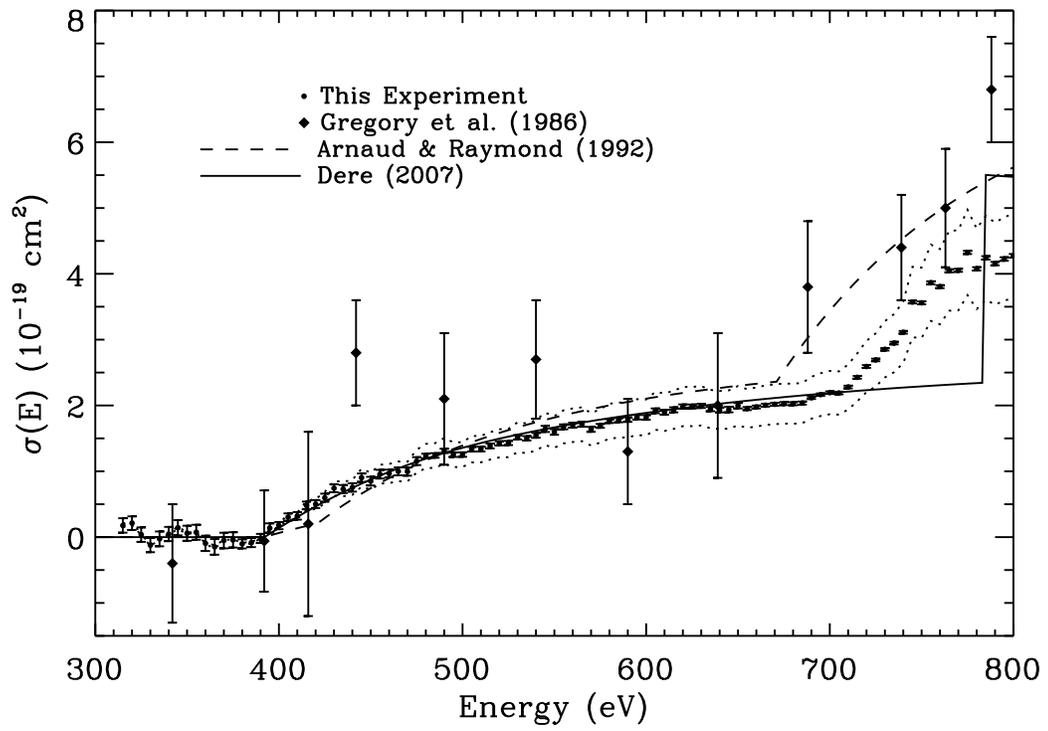}
\caption{\label{fig:fe13thresh} A portion of Figure~\ref{fig:fe13cross} focussing on the threshold energy range. 
}
\end{figure}

	Excitation of a $3s$ electron to an autoionizing level is not included in these theoretical cross sections. Using the LANL Atomic Code \citep{Magee:ASP:1995}, we have estimated that EA is possible for $3\rightarrow n \gtrsim 10$ excitations and could contribute to the ionization cross section near threshold (Figure~\ref{fig:fe13thresh}). Previous measurements for other $3s^2\,3p^q$ ions have found a significant EA contribution via this channel ($q=2$, \citealt{Hahn:ApJ:2011a}; $q=3$, \citealt{Hahn:ApJ:2011}; $q=4$~and~5, \citealt[][]{Hahn:ApJ:2012}) which has been confirmed by theoretical calculations ($q=3$, \citealt{Kwon:PRA:2012}). However, here we find little discrepancy between theory and experiment near threshold, despite the omission of $3\rightarrow n$ EA in the calculations. The $n$ required for the excitation to autoionize is about the same for $\fethirteen$ as for the ions of previous measurements, for example for Fe$^{11+}$ $3\rightarrow n > 8$ excitations were autoionizing. The reason this EA channel is relatively smaller here could be due to there being only one $3p$ electron. The effect could also be masked by the 16\% systematic uncertainty. 
	
	At higher energies there are some discrepancies in the shape of the cross section. The \citet{Dere:AA:2007} calculation overestimates the magnitude of $2\rightarrow4$~EA at about 1020~eV. This discrepancy has been found previously for similar ions \citep{Hahn:ApJ:2011, Hahn:ApJ:2011a, Hahn:ApJ:2012}. A possible explanation for the discrepancy is that the calculations underestimate the branching ratios for radiative stabilization. Another possibility is that the calculations underestimate the branching ratio for auto-double ionization \citep{Kwon:PRA:2012}, but this explanation is less likely since we do not observe any corresponding increase in the EIDI cross section at 1020~eV (see Section~\ref{subsec:double}). 
	
\begin{figure}
\centering \includegraphics[width=0.9\textwidth]{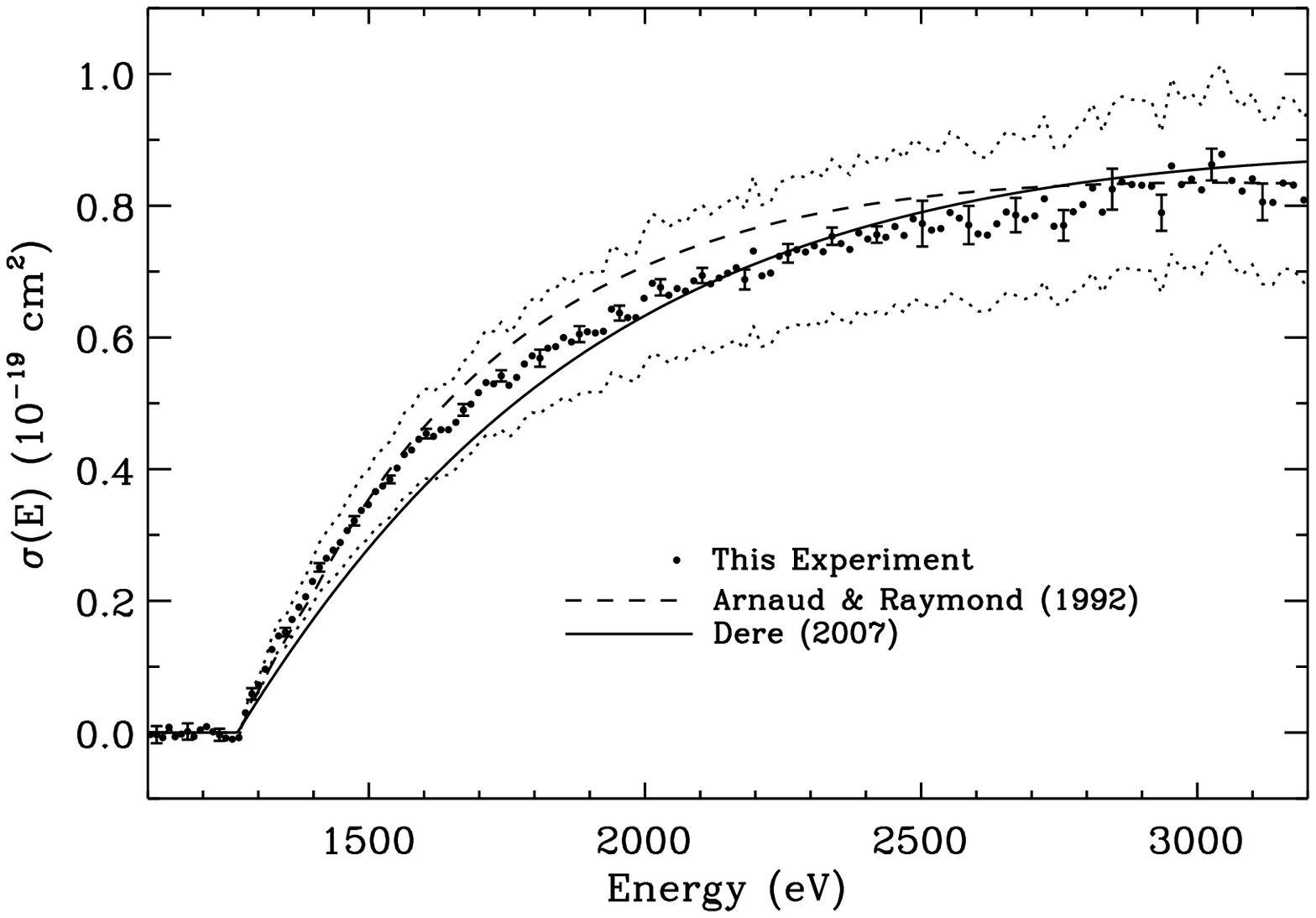}
\caption{\label{fig:fe16cross} Same as Figure~\ref{fig:fe13cross}, but for $\fesixteen$ forming Fe$^{17+}$.
}
\end{figure}

\begin{figure}
\centering \includegraphics[width=0.9\textwidth]{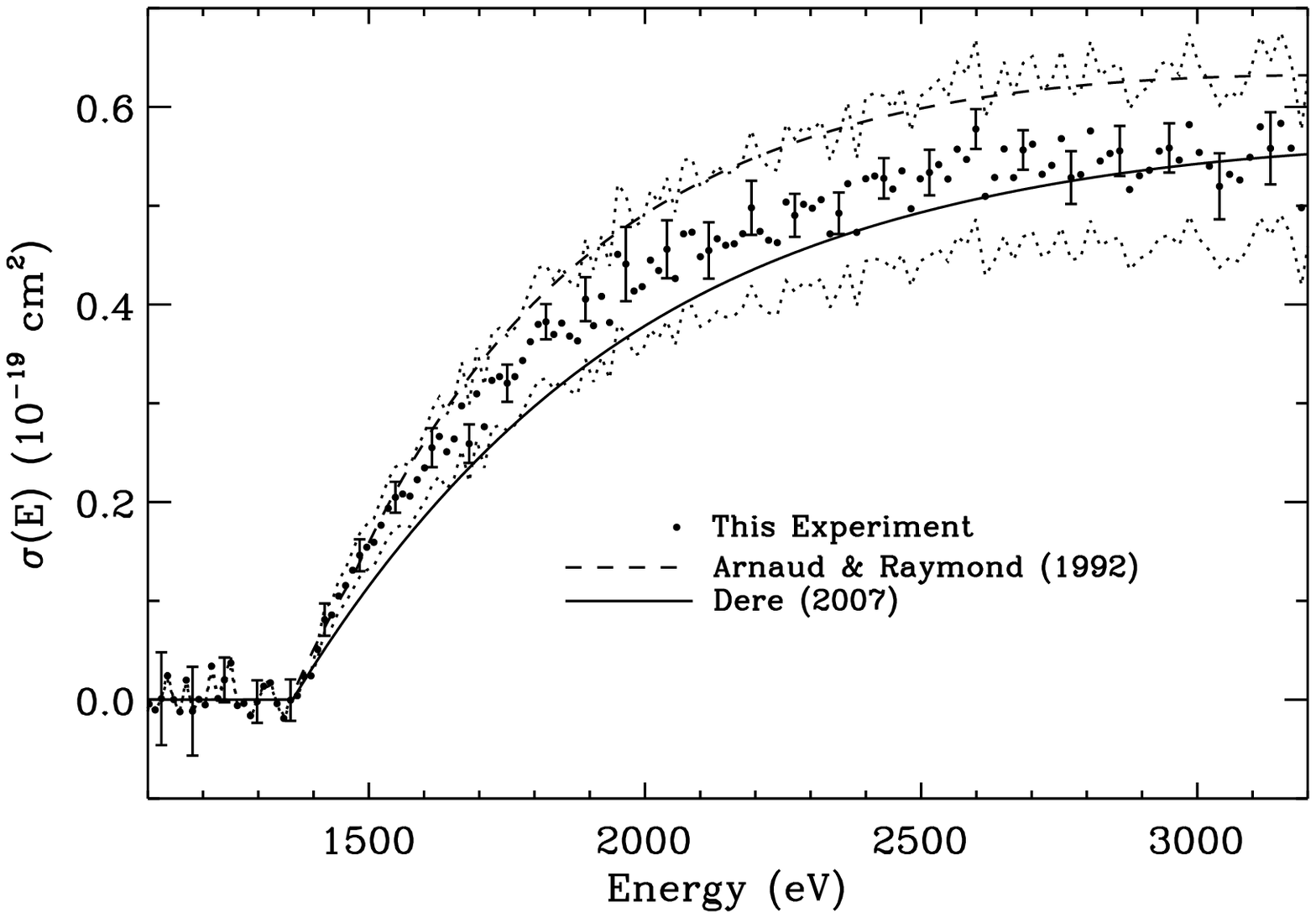}
\caption{\label{fig:fe17cross} Same as Figure~\ref{fig:fe16cross}, but for $\feseventeen$ forming Fe$^{18+}$. 
}
\end{figure}

	Figures~\ref{fig:fe16cross} and \ref{fig:fe17cross} show the measured EISI cross sections for $\fesixteen$ forming Fe$^{17+}$ and $\feseventeen$ forming Fe$^{18+}$, respectively. These data are available in the electronic edition of this journal as tables following the format of Tables~\ref{table:fe16cross} and \ref{table:fe17cross}. The figures also illustrate the cross sections recommended by \citet{Arnaud:ApJ:1992}, which is based on the calculations of \citet{Younger:JQSRT:1982}, and the theoretical cross section of \citet{Dere:AA:2007}. For each ion there is generally good agreement with the measurement, to within the experimental uncertainties. 
		
	One discrepancy between experiment and theory is that the measured cross sections for both ions increase faster close to threshold than predicted by \citet{Dere:AA:2007}. This may be due to EA from excitation of a $2s$ electron to an energy above the threshold for ionization of the $2p$ electron. It should be noted, though, that the cross section of \citet{Arnaud:ApJ:1992} agrees very well with our results near threshold despite being based on calculations that included only direct ionization.
	
\subsubsection{Rate Coefficients} \label{subsubsec:rate}

\begin{figure}
\centering \includegraphics[width=0.9\textwidth]{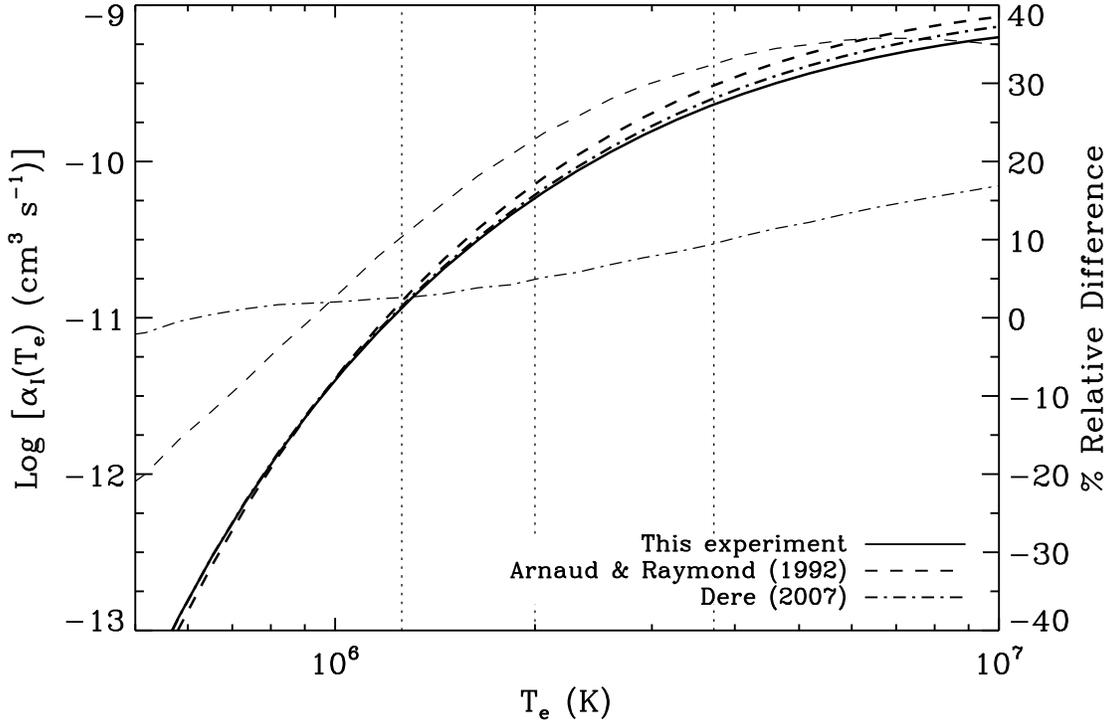}
\caption{\label{fig:fe13rate} Thick lines show various plasma rate coefficients for $\fethirteen$ forming Fe$^{14+}$. The solid curve indicates the experimental results, which can be read off the left axis. They are compared to the theoretical results of \citet[][dashed curve]{Arnaud:ApJ:1992} and \citet[][dash-dotted curve]{Dere:AA:2007}. The relative difference between these and the present results, (theory-experiment)/experiment, are shown by thin lines, with values read off the right axis. The dotted vertical lines denote the temperature range where $\fethirteen$ is $>1\%$ abundant in collisional ionization equilibrium with the center line at the temperature of peak $\fethirteen$ abundance \citep{Bryans:ApJ:2009}. 
}
\end{figure}

\begin{figure}
\centering \includegraphics[width=0.9\textwidth]{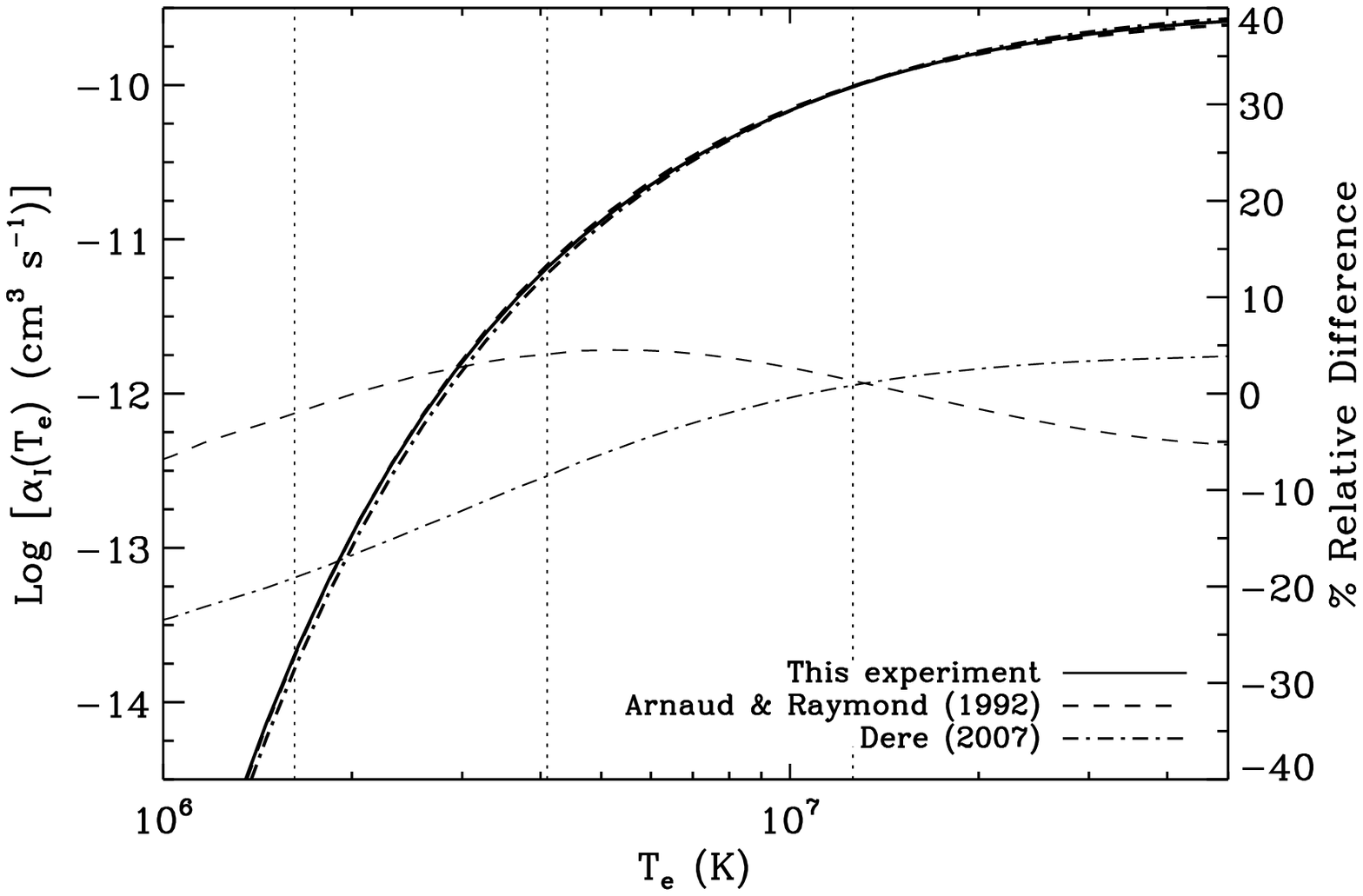}
\caption{\label{fig:fe16rate} Same as Figure~\ref{fig:fe13rate}, but for $\fesixteen$ forming Fe$^{17+}$. 
}
\end{figure}

\begin{figure}
\centering \includegraphics[width=0.9\textwidth]{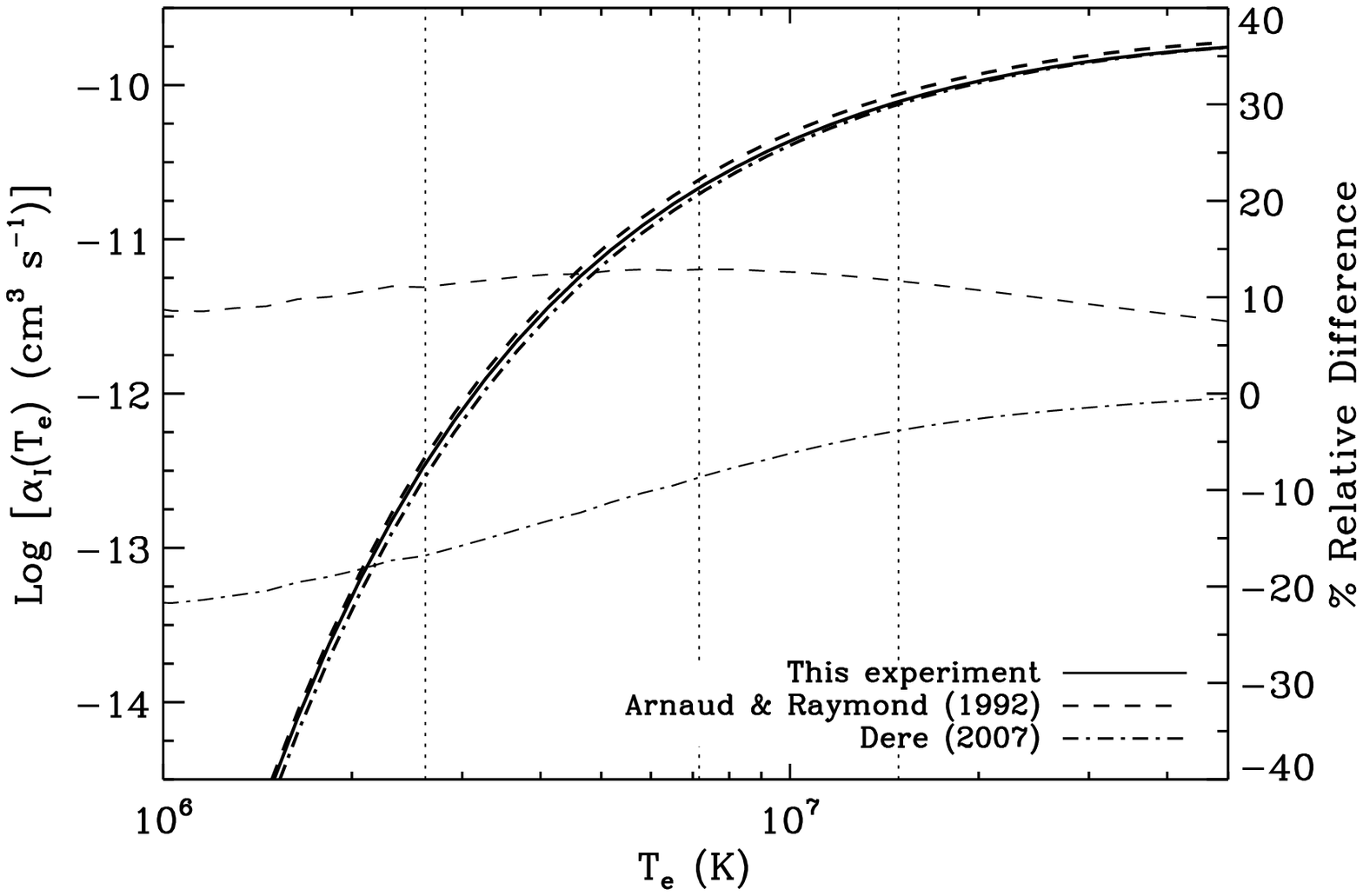}
\caption{\label{fig:fe17rate} Same as Figure~\ref{fig:fe13rate}, but for $\feseventeen$ forming Fe$^{18+}$. 
}
\end{figure}

	Using the measured cross sections, we have derived EISI plasma ionization rate coefficients $\alpha_{\mathrm{I}}$ as a function of electron temperature $T_{\mathrm{e}}$ \citep[cf.,][]{Hahn:ApJ:2011}. Figures~\ref{fig:fe13rate}, \ref{fig:fe16rate}, and \ref{fig:fe17rate} show the results for $\fethirteen$, $\fesixteen$, and $\feseventeen$, respectively. In each case we compare these results to the rate coefficients of \citet{Arnaud:ApJ:1992} and \citet{Dere:AA:2007}. The vertical dotted lines in the figures indicate for ionization equilibrium the temperature ranges over which each ion is more than 1\% abundant relative to the total Fe abundance as well as the temperature of peak abundance \citep{Bryans:ApJ:2009}. 
	
		$\fethirteen$ is abundant from $1.3\times10^{6}$~K to $3.7\times10^6$~K with a peak abundance at $2.0\times10^{6}$~K. In this range the $\fethirteen$ rate coefficients of \citet{Arnaud:ApJ:1992} differ from our experiment by up to $32\%$, while those of \citet{Dere:AA:2007} agree with our results to within $10\%$. $\fesixteen$ is abundant from $1.6\times10^6$~K to $1.3\times10^7$~K peaking at $4.1\times10^6$~K. Surprisingly, the older rate coefficients of \citet{Arnaud:ApJ:1992} agree with our results to within 4\% within this range. The more recent data of \citet{Dere:AA:2007} differ by 19\% at the low end of the temperature range due to the discrepancy near the ionization threshold. At the temperature of peak abundance they differ by $9\%$ and at the high $T_{\mathrm{e}}$ end they agree with our measurement to within 1\%. Finally, $\feseventeen$ is abundant from $2.6\times10^6$~K to $1.5\times10^7$~K with peak abundance at $7.2\times10^6$~K. The rate coefficients of \citet{Arnaud:ApJ:1992} agree with our measurements in this range to within 13\%. Those of \citet{Dere:AA:2007} differ from the experimentally derived result by 17\% at the low end of the $T_{\mathrm{e}}$ range (for the same reasons as for Fe$^{16+}$), by $9\%$ at peak abundance, and by only 4\% at the high $T_{\mathrm{e}}$ end. 
	
	The energy range covered by the experiment does not include excitation or ionization of a $1s$ electron. These channels are also neglected by \citet{Arnaud:ApJ:1992} and \citet{Dere:AA:2007}. This neglect may introduce a small error in the calculation of the ionization rate coefficient for $\fesixteen$ and $\feseventeen$. The reason is that in deriving $\alpha_{\mathrm{I}}(T_{\mathrm{e}})$, the integration over the cross section is performed up to $E_{0} + 6k_{\mathrm{B}}T_{\mathrm{e}}$, where $E_0$ is the ionization threshold and $k_{\mathrm{B}}$ is the Boltzmann constant \citep{Fogle:ApJS:2008}. Thus, to calculate the rate coefficients over the full range where these ions are abundant, the integration should be performed up to 7984~eV for $\fesixteen$ and up to 9113~eV for $\feseventeen$. These are greater than the measured energy range, and we have extrapolated the measurements to higher energies by scaling the \citet{Dere:AA:2007} cross section to our measurements. This leaves an uncertainty as the integration limits exceed the $1s$ excitation and ionization thresholds. For $\fesixteen$ the lowest EA channel is the $1\rightarrow3$~EA, which opens at $\approx 7150$~eV. For $\feseventeen$ the lowest channel is $1\rightarrow2$~EA, which opens at $\approx6450$~eV \citep{Hou:ADNDT:2009,NIST:2012}. The threshold for $1s$ ionization is 7714.7~eV for $\fesixteen$ and 7823.2~eV for $\feseventeen$ \citep{NIST:2012}. 
	
	In order to assess the possible error from neglecting direct ionization and EA we have estimated the $1s$ ionization and excitation cross sections using the LANL Atomic Physics Code \citep{Magee:ASP:1995}. These calculations show that the $1s$ direct ionization cross section is $\sim 0.5\%$ of the $n=2$ direct ionization cross section for these ions in the relevant energy ranges. The maximum $1s$ EA cross section is the total excitation cross section. At the relevant energies, compared to the included EISI from $n=2$, the $\fesixteen$ $1\rightarrow3$ excitation cross section is $\sim 0.05\%$ and the $\feseventeen$ $1\rightarrow2$ excitation cross section is about $0.5\%$. The contribution to single ionization from $n=1$ excitation and ionization is actually smaller than implied by these cross sections. In the case of $1\rightarrow n$ EA, the continuum state is estimated to radiatively stabilize 40\% of the time and lead to EISI only for the other 60\% \citep{Kaastra:AAS:1993}. For $1s$ ionization, radiative relaxation of the intermediate state completes the EISI process 40\%, but the system autoionizes the other 60\% leading to EIDI. Given the small cross sections for $1\rightarrow n$ EA and $1s$ ionization, the branching ratios of the intermediate states, and the small fraction of the integrated energy range where they contribute at all, we expect that the omission of these processes has negligible effect on the calculated rate coefficients over the temperature ranges where $\fesixteen$ and $\feseventeen$ are abundant. 

	Table~\ref{table:coeff} presents coefficients for a polynomial fit to the scaled rate coefficient $\rho(x)=10^{-6}\sum_{i}{a_i x^i}$, which can be used to reproduce the plasma rate coefficients. The rate coefficient $\alpha_{\mathrm{I}}(T_{\mathrm{e}})$ is related to the scaled rate coefficient $\rho$ by \citep{Dere:AA:2007}: 
\begin{equation}
\alpha_{\mathrm{I}}(T_{\mathrm{e}}) = t^{-1/2}E_0^{-3/2}E_{1}(1/t)\rho(x), 
\label{eq:invscalerate}
\end{equation}
where $E_{1}(1/t)$ is the first exponential integral and $t=k_{\mathrm{B}}T_{\mathrm{e}}/ E_0$ with $E_0$ the ionization threshold (392.2~eV for $\fethirteen$, 1262.7 for $\fesixteen$, and 1357.8 for $\feseventeen$). The scaled temperature $x$ is given by
\begin{equation}
x = 1 - \frac{\ln 2}{\ln(t+2)}
\label{eq:invx}
\end{equation}
and by inverting $T_{\mathrm{e}}$ can be obtained from $x$: 
\begin{equation}
T_{\mathrm{e}} = \frac{E_0}{k_{\mathrm{B}}}\left[\exp\left(\frac{\ln 2}{1-x} \right) - 2 \right]. 
\label{eq:invscaletemp}
\end{equation}
The experimental rate coefficents are reproduced to 1\% accuracy or better for $T_{\mathrm{e}} = 4\times10^{5}$ -- $1\times10^{8}$~K for $\fethirteen$, $T_{\mathrm{e}} = 6\times10^{5}$ -- $1\times10^{8}$~K for $\fesixteen$ and $T_{\mathrm{e}} = 1.5\times10^{6}$ -- $1\times10^{8}$~K for $\feseventeen$.  	

\subsection{Double Ionization}\label{subsec:double}

\begin{figure}
\centering \includegraphics[width=0.9\textwidth]{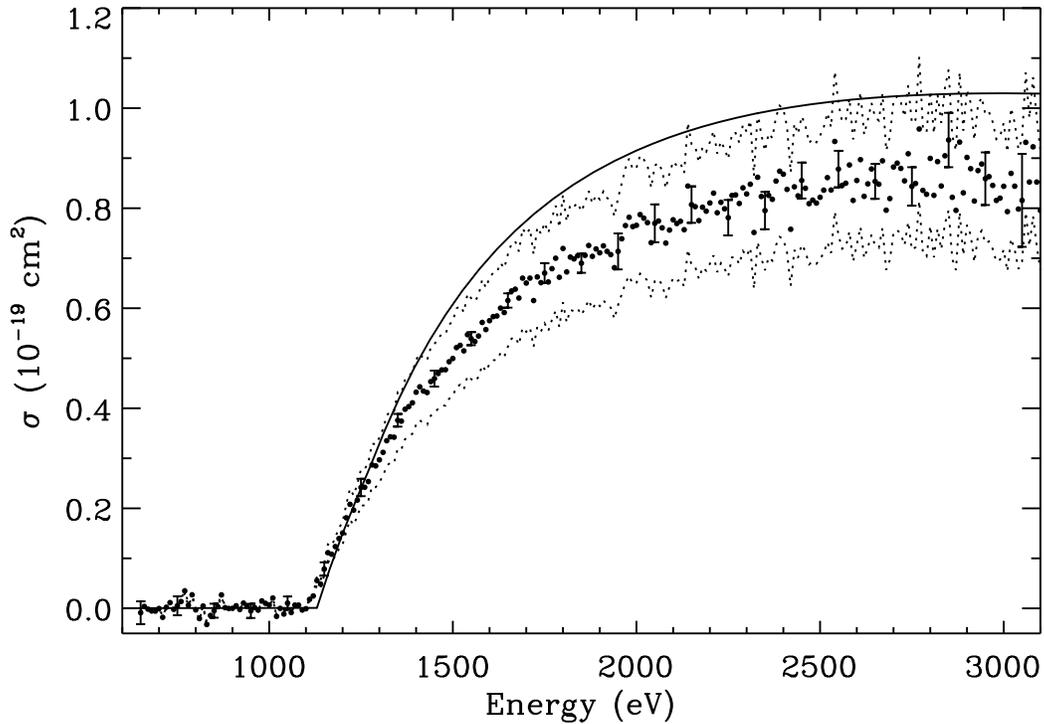}
\caption{\label{fig:fe13di} EIDI cross section for $\fethirteen$ forming Fe$^{15+}$. The statistical uncertainty is indicated by error bars on selected points and the systematic uncertainties are illustrated by the dotted curves. The solid curve shows an estimate for the EIDI cross section due to direct ionization of an $n=2$ electron forming a state that relaxes by emission of a second electron producing, in total, double ionization (see the text). 
}
\end{figure}

	Figure~\ref{fig:fe13di} shows the measured $\fethirteen$ double ionization cross section. The dotted curves give the systematic uncertainty and the error bars illustrate the statistical uncertainty for select points. These data are available in the electronic edition of this journal as a table following the format of Table~\ref{table:fe13di}. Although the threshold for double ionization is 848.4~eV, the cross section is consistent with zero until about 1100~eV when ionization of an $n=2$ electron becomes possible. The solid line illustrates the expected double ionization cross section due to ionization of an $n=2$ electron forming a state that relaxes through autoionization. This cross section was calculated using the LANL Atomic Physics Code \citep{Magee:ASP:1995} to determine the cross section for single ionization of an $n=2$ electron and then scaling the result by the Auger yields of about 93\% for $2p$ ionization and 95\% for $2s$ ionization given by \citet{Kaastra:AAS:1993}. In the measured energy range, essentially all of the double ionization of $\fethirteen$ is due to this process. This is consistent with results from other highly charged ions \citep{Muller:PRL:1980,Muller:JPhysB:1985,Stenke:JPhysB:1999d,Hahn:ApJ:2011, Hahn:ApJ:2011a}. 
	
\section{Summary}\label{sec:sum}	

	We have measured cross sections for EISI from the ground states of $\fethirteen$, $\fesixteen$, and $\feseventeen$. For $\fethirteen$ we find discrepancies of about 40\% when compared to an earlier crossed beams experiment. This is likely due to metastable ions in that work and their absence in ours. The theoretical cross section recommended by \citet{Arnaud:ApJ:1992} is more than 30\% larger than our result. The recent calculation of \citet{Dere:AA:2007} is also $20\%$ larger than our result. The discrepancy with these theoretical calculations appears to be due to their treatment of EA. In particular, we do not observe the contribution from $2\rightarrow4$ EA predicted by \citet{Dere:AA:2007}. For $\fesixteen$ and $\feseventeen$ our results generally agree with theory to within the experimental uncertainties. There is a small discrepancy in that the experimental cross section rises faster near threshold than predicted by \citet{Dere:AA:2007}. One possibility is that this is due to neglecting EA from $n=2$ excitations in the calculations. The measured EIDI cross section for $\fethirteen$ is dominated by $n=2$ ionization followed by autoionization of the excited state leading to a net double ionization. This result is consistent with measurements for other ions. 
	
\begin{acknowledgments}
	We appreciate the efficient support by the MPIK accelerator and TSR groups during the beamtime. This work was supported in part by the NASA Astronomy and Physics Research and Analysis program and the NASA Solar Heliospheric Physics program. We also acknowledge financial support by the Max Planck Society, Germany and from Deutsche Forschungsgemeinschaft (contract no. Schi 378/8-1).  
\end{acknowledgments}

\begin{deluxetable}{lccccc}
\tablewidth{0pt}
\tablecaption{Sources of Uncertainty.
\label{table:err}}
\tablehead{
	\colhead{} & 
	\colhead{Source} & 
	\multicolumn{4}{c}{Estimated $1\sigma$ Uncertainty} \\
	
	\colhead{} & 
	\colhead{} & 
	\multicolumn{2}{c}{\hspace{30pt}$\fethirteen$} & 
	\colhead{$\fesixteen$} & 
	\colhead{$\feseventeen$} \\
	
	\colhead{} & 
	\colhead{} & 
	\colhead{EISI} & 
	\colhead{EIDI} & 
	\colhead{EISI} &
	\colhead{EISI} 	
	}
\startdata
&Counting statistics 										& 2\% 					& 3\% 	& 3\% 			& 6\% 			\\
&Detector efficiency										& 3\%						& 3\% 	& 3\% 			& 3\% 			\\
&Ion current measurement 								& 15\% 					& 15\% 	& 15\%			& 15\% 			\\
&Electron density 											& 3\% 					& 3\% 	& 3\%				& 3\%				\\
&Pressure fluctuations\tablenotemark{1} & 0.2\% (0.8\%) & --		& 2\% (4\%)	& 2\% (5\%)	\\
\hline
&Quadrature sum 												& 16\% (16\%)		& 16\% 	& 16\% (16\%)& 17\% (17\%) \\
\enddata
\tablenotetext{1}{The uncertainties in parentheses refer to the energy range where the reference point was above the ionization threshold. This is $>1050$~eV for $\fethirteen$ EISI, $>1500$~eV for $\fethirteen$ EIDI, $>1980$~eV for $\fesixteen$ EISI, and $>1950$~eV for $\feseventeen$ EISI. For $\fethirteen$ EIDI no pressure fluctuations were observed.}
\end{deluxetable}

\clearpage

\begin{deluxetable}{lll}
\tablewidth{0pc}
\tablecaption{$\fethirteen$ Single Ionization Cross Section.
\label{table:fe13cross}}
\tablehead{
	\colhead{$E$~(eV)} & 
	\colhead{$\sigma_{\mathrm{I}}$~(cm$^2$)} &
	\colhead{Statistical Error}
}
\startdata
400 & 1.7490E-20 & 5.4474E-21 \\
550 & 1.5980E-19 & 4.2844E-21 \\
700 & 2.1955E-19 & 1.9516E-21 \\
850 & 4.6733E-19 & 2.7333E-21 \\
1000 & 4.8221E-19 & 2.4091E-21 \\
1500 & 4.1659E-19 & 2.7346E-21 \\
2000 & 3.6449E-19 & 1.6042E-20
\enddata
\tablecomments{There is a systematic uncertainty of 16\% in the cross section (see the text). Table~\ref{table:fe13cross} is published in its entirety in the electronic edition of this journal.}
\end{deluxetable}

\begin{deluxetable}{lll}
\tablewidth{0pc}
\tablecaption{$\fesixteen$ Single Ionization Cross Section.
\label{table:fe16cross}}
\tablehead{
	\colhead{$E$~(eV)} & 
	\colhead{$\sigma_{\mathrm{I}}$~(cm$^2$)} &
	\colhead{Statistical Error}
}
\startdata
1300.5 & 7.2389E-21 & 8.8629E-22 \\
1499.3 & 3.4611E-20 & 6.0735E-22 \\
1998.4 & 6.5945E-20 & 1.2343E-21 \\
2502.2 & 7.7283E-20 & 3.4703E-21 \\
3007.8 & 8.2407E-20 & 2.4911E-21
\enddata
\tablecomments{There is a systematic uncertainty of 16\% in the cross section (see the text). Table~\ref{table:fe16cross} is published in its entirety in the electronic edition of this journal.}
\end{deluxetable}

\begin{deluxetable}{lll}
\tablewidth{0pc}
\tablecaption{$\feseventeen$ Single Ionization Cross Section.
\label{table:fe17cross}}
\tablehead{
	\colhead{$E$~(eV)} & 
	\colhead{$\sigma_{\mathrm{I}}$~(cm$^2$)} &
	\colhead{Statistical Error}
}
\startdata
1407.7 & 5.0680E-21 & 1.6703E-21 \\
1601.0 & 2.3464E-20 & 1.5598E-21 \\
1995.0 & 4.1807E-20 & 3.7005E-21 \\
2498.4 & 5.2723E-20 & 2.5672E-21 \\
3003.6 & 5.5388E-20 & 2.5631E-21
\enddata
\tablecomments{There is a systematic uncertainty of 16\% in the cross section (see the text). Table~\ref{table:fe17cross} is published in its entirety in the electronic edition of this journal.}
\end{deluxetable}

\begin{deluxetable}{llll}
\tablewidth{0pc}
\tablecaption{Fifth-order Polynomial Fitting Parameters to Reproduce the Scaled Single Ionization Rate Coefficient $\rho = 10^{-6}\sum_{i=0}^{i=5}{a_{i}x^{i}}$~$\mathrm{cm^{3}\,s^{-1}\,eV^{3/2}}$ (see equations \ref{eq:invscalerate} and \ref{eq:invscaletemp}).
\label{table:coeff}}
\tablehead{
	
	\colhead{$i$} & 
	\multicolumn{3}{c}{$a_{i}$} \\

	\colhead{} & 
	\colhead{$\fethirteen$} &
	\colhead{$\fesixteen$} & 
	\colhead{$\feseventeen$} 
}
\startdata 
0 & \phs9.59988 		& \phs27.1411 		& \phs17.4957 \\
1 & \phn-53.2715 		& \phn-24.4446		& \phs63.0820 \\
2 & \phs401.820 		& \phs43.0958			& \phn-485.718 \\
3 & \phn-1018.29		& \phs74.8937 		& \phs1517.84 \\
4 & \phs1115.84 		& \phn-252.597 		& \phn-2127.02 \\
5 & \phn-455.515 		& \phs150.090			& \phs1094.75
\enddata
\end{deluxetable}

\begin{deluxetable}{lll}
\tablewidth{0pc}
\tablecaption{$\fethirteen$ Double Ionization Cross Section.
\label{table:fe13di}}
\tablehead{
	\colhead{$E$~(eV)} & 
	\colhead{$\sigma_{\mathrm{I}}$~(cm$^2$)} &
	\colhead{Statistical Error}
}
\startdata
900 & 5.5118E-24 & 1.3882E-21 \\
1100 & -7.6013E-23 & 1.2907E-21 \\
1300 & 2.9696E-20 & 1.2073E-21 \\
1500 & 4.9954E-20 & 1.2722E-21 \\
2000 & 7.6579E-20 & 5.1443E-21 \\
2500 & 8.2175E-20 & 3.5210E-21 \\
3000 & 8.4374E-20 & 4.9785E-21 
\enddata
\tablecomments{There is a systematic uncertainty of 16\% in the cross section (see the text). Table~\ref{table:fe13di} is published in its entirety in the electronic edition of this journal.}
\end{deluxetable}

\bibliography{TSR_Bib}

\end{document}